# Gossip-based Signaling Dissemination Extension for Next Steps In Signaling

M. Femminella, R. Francescangeli, G. Reali
DIEI – University of Perugia
Perugia, Italy
{femminella,francescangeli,reali}@diei.unipg.it

H. Schulzrinne
CS – Columbia University
New York, USA
hgs@cs.columbia.edu

*Abstract*— **In this paper, we propose a new gossip-based signaling dissemination method for the Next Steps in Signaling protocol family. In more detail, we propose to extend the General Internet Signaling Transport (GIST) protocol, so as to leverage these new dissemination capabilities from all NSIS Signaling Layer Protocol applications using its transport capabilities. The new GIST extension consists of two main procedures: a bootstrap procedure, during which new GIST-enabled nodes discover each other, and a service dissemination procedure, which is used to effectively disseminate signaling messages within an Autonomous System. To this aim, we defined three dissemination models, bubble, balloon, and hose, so as to fulfill requirements of different network and/or service management scenarios. An experimental campaign carried out on the GENI testbed shows the effectiveness of the proposed solution.**

*Keywords: Epidemic; GIST; NSIS; gossip; signaling; routing*

## I. INTRODUCTION

The Internet is continuously changing and expanding over time, open to the introduction of any new type of network, with undefined space and service borders, creating more and more challenges for network operators regarding its operation and management. A basic constituent of the ever-evolving current and future Internet is the set of signaling protocols used to maintain, monitor or change the state of nodes in most network types. The majority of current signaling protocols are designed to function as a specific application protocol, offering a closed set of well-defined services. Consequently, every time a new application is conceived, a new signaling protocol, that re-engineers some of the same basic functionalities required for signaling, is likely to be created.

In order to overcome the perceived shortcomings of this approach, the IETF Next Steps in Signaling (NSIS) [1] Working Group was formed in November 2001. The goal of the NSIS project was to design and develop a new generic Internet signaling protocol suite, whose first use case focused on Quality of Service (QoS) [2] signaling, interoperability and mobility. Specifically, NSIS first tried to overcome known issues of the Resource ReSerVation Protocol (RSVP), taking it as the background for the development work, and then added middlebox and NAT traversal as a second use case [3]. The development of a generic IP signaling protocol which meets various signaling objectives constituted a flexible and an effective solution in terms of development effort, signaling compatibility, and operation/management costs.

The resulting NSIS framework has been designed to be modular and easily extendable in all its entities, while trying to be secure, lightweight and generic enough to accommodate all kinds of signaling application needs. However, the main NSIS shortcoming is that, being primarily designed based on RSVP and being its QoS-aware nodes discovery process stuck to the data path, the NSIS-based signaling can only install, modify and remove state either on two end hosts communicating with each other (end-to-end signaling), or on a set of intermediate nodes on the data-path between two end hosts (path-coupled signaling). This is typically achieved by using end-to-end addressed messages equipped with an IP Router Alert Option (RAO) that are routed along the data-path and intercepted at NSIS-capable nodes. Thankfully, the NSIS architecture has been designed to be flexible [4], and the routing of signaling messages is controlled by the so-called Message Routing Method (MRM) that is applied to each NSIS message. By creating a new MRM, not only signaling messages can be routed to different NSIS nodes with respect to those lying on the data path between two end-hosts, but messages can even become completely decoupled from the idea of an end-to-end network path, eventually resulting in no direct relation to any underlying data flow. By extending the layered architecture of NSIS with new MRMs, a whole new range of signaling applications becomes immediately available to all nodes that implement the new feature without having to modify the logic of the application itself. Applications that immediately benefit from such extensions include and are not limited to fixed and mobile overlay networks setup and management [24], in-network services installation [22][26][27], advanced media and cache services [7], cloud management, and all services falling into the new paradigm of Software-Defined Networking [25].

In this paper we present the design and implementation of a new gossip-based, or epidemic, signaling MRM for the General Internet Signaling Transport (GIST) protocol [5]. GIST is the most commonly used implementation of the NSIS Transport Layer Protocol, thus by extending GIST our new method indirectly extends the routing and signaling capabilities of nearly all existing NSIS compatible nodes. Also, since NSIS nodes are incrementally deployed in real networks, they tend to be scattered around the autonomous system (AS) and for this reason they can be regarded as an overlay network. In order to effectively propagate signaling messages using the presented epidemic MRM through a NSIS overlay network, an efficient NSIS node discovery approach is then required. To this aim, in this paper we also introduce a new efficient gossip-based NSIS node discovery process that exploits the native packet interception capabilities of GIST. This new approach is used to bootstrap any NSIS network by collecting neighbor NSIS nodes capabilities and calculating various metrics subsequently employed to control the epidemic signaling message delivery. In order to evaluate the effectiveness and efficiency of our proposed solutions, we used a NSIS network scenario based on the well-known NSF network topology implemented on the GENI [6] virtual laboratory framework. The tests have been performed by triggering off-path installation of a network monitoring and reporting service in all compatible nodes of an Autonomous System (AS) by using NetServ [7][22] and NSIS-ka [8], an open source implementation of NSIS.

The paper is organized as follows. Section II describes the NSIS architecture and how it can be extended, focusing in particular on the GIST layer. In section III, we present our gossip-based NSIS node discovery process and the proposed dissemination MRM. Section IV contains our experimental results together with the GENI testbed description. Section V discusses about related work and possible alternatives to our epidemic MRM. Finally, section VI draws our conclusions and lists our future work activities.

## II. BACKGROUND: THE NSIS PROTOCOL FAMILY

The NSIS protocol suite essentially divides the signaling functionality into two layers as shown in Figure 1. The lower layer, the NSIS Transport Layer Protocol (NTLP), is in charge of transporting the higher-layer protocol messages to the next signaling node on path. This includes discovery of the next-hop NSIS node, which may not be the next routing hop, and different transport and security services depending on the signaling application requirements. The actual signaling application logic is implemented in the higher layer of the NSIS stack, the NSIS Signaling Layer Protocol (NSLP). The NSIS protocol suite supports both IPv4 and IPv6.

GIST has been developed as the protocol that fulfills the role of the NTLP. Thus, it provides signaling transport and security services to NSLPs and the associated signaling applications. GIST does not define new IP transport protocols or security mechanisms but rather makes use of existing protocols, such as TCP, UDP, TLS, and IPsec. Applications can specify the desired transport attributes for the signaling flow, e.g., unreliable or reliable, and GIST then chooses the most appropriate transport protocol(s) to satisfy the requirements of the flow.

GIST is only concerned with transporting NSLP messages hop-by-hop between pairs of signaling nodes while the end-to-end signaling functionality, if needed, is provided by the upper layer NSLPs. Messages transmitted by GIST on behalf of an NSLP are identified, among the other things, by a unique NSLP identifier (NSLPID), assigned by IANA to the NSLP. Only two NSLP protocols are currently specified as IETF RFCs (Quality-of-Service signaling [2] and NAT/firewall traversal [3]) but several others have been designed and implemented on top of NSIS by the community, an example is the signaling protocol of the NetServ [7][22] architecture.

Being based primarily on RSVP, NSIS has been primarily designed to provide the signaling needed to install state on nodes that lie on the path that will be taken by some end-to-end flow of data packets. This is achieved by routing signaling messages along the same path and intercepting the signaling packet at NSIS-capable nodes. However, the NSIS architecture, and in particular the GIST protocol, is designed to be flexible and extendable, and the routing of signaling messages is controlled by the Message Routing Method (MRM) that is applied to each one of them. A MRM is defined as the algorithm used by GIST to discover NSIS nodes and route signaling messages. The initial GIST specifications define:

- the Path Coupled MRM designed to drive signaling along the path that will be followed by the data flow;
- the Loose End MRM, used for preconditioning the state in firewalls and NAT when data flow destinations lie behind this sort of middleboxes.

Parameters carried by each signaling message drive the operation of the relevant transport or signaling application. In particular, each GIST message carries a Message Routing Information (MRI) object that allows NSIS nodes to identify the MRM used and to route signaling messages accordingly.

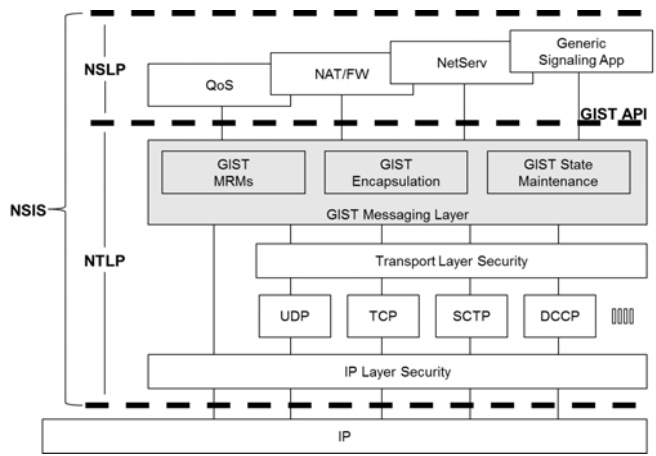

Figure 1 – NSIS framework: 2-layers architecture with detailed transport and supported GIST entities.

When the NSIS WG started working on a generic and layered IP signaling architecture, it tried to contain the complexity of the task by imposing two main restrictions: (i) focus on path-coupled signaling, and (ii) no multicast support. These restrictions can be lifted thanks to the modular architecture of NSIS and GIST, particularly thanks to their extensibility in supporting different MRMs. In fact, as specified in [5] and [4], the GIST design can be extended to cater for multicast flows and for situations where signaling is not tied to an end-to-end data flow, as in our epidemic case. For example, one possible additional MRM under development is documented in [20]. This MRM targets signaling messages towards an explicit target address other than the data flow destination and is intended to assist handover operations.

## III. EPIDEMIC DISSEMINATION EXTENSION

In this section we present the design and implementation of the gossip-based GIST node discovery process, together with the MRM signaling dissemination extension of GIST.

*A. GIST node discovery process*

In order to efficiently implement a gossip-based signaling dissemination extension, every GIST node has to be aware of the presence of other GIST nodes in the network and has to calculate some utility metrics. To actually communicate, a node has to know the IP addresses of the other nodes.

Then, first of all, all GIST nodes have to perform a bootstrap procedure, followed by the actual node discovery phase. When a node first boots up it has no knowledge of other GIST nodes and thus it is not able to contact them directly. In order to expand its knowledge of the network it needs to contact a bootstrapping node. This is a GIST node that provides initial information to newly joining nodes so that they may successfully join the overlay network and share information to perform epidemic-style signaling dissemination. A newly joining node may identify which nodes are bootstrapping nodes in different ways. New nodes may be pre-configured with static addresses of the bootstrapping nodes. In such a case, the bootstrapping node addresses cannot change, and therefore they should be stable members of the network, i.e. fault-tolerant and with no desire to leave the network. Alternatively, the bootstrap node can be identified via a DNS service, where a domain name resolves to one of the bootstrapping nodes' addresses. This allows the bootstrapping nodes' addresses to change as needed, and overlay stable nodes are therefore not necessary. To allow for an easier deployment of our gossip-based dissemination extension, we designed a completely automatic GIST node configuration procedure, based on the second option. The only requirement in our case is that each GIST node, when first booting up, has to know its own IP address, which is clearly reasonable. From the IP address, the node finds out the number of the AS (ASN) to which it belongs to, by doing a whois query to an IP-to-ASN mapping public service, such as that provided by Team Cymru Research NFP [21]. An example, using an IP address of a router currently in use at the University of Perugia, is shown in Figure 2. As soon as the ASN is obtained, the GIST node performs a DNS query about the domain name obtained by combining together the ASN and the generic "nsis.org" domain. For instance, after the previous query shown in Figure 2, the node would make a DNS query to resolve the *as137.nsis.org* name. The DNS query will then return one or more IP addresses belonging to bootstrapping nodes and thus completing the automatic configuration procedure. Clearly, in order to work, our procedure needs the creation of a generic *container* domain, such as "nsis.org" to which AS operators can register the IP addresses of the bootstrapping nodes of their own networks. The final architecture we envision is very similar to the one used by the NTP pool project [14] that provides NTP service for millions of clients around the world. The difference is in the domain name hierarchy creation, that in the case of NTP pool is geographically based, whereas in our case is based on the ASN.

```
$ whois -h whois.cymru.com " -v 141.250.40.34"

[Querying v4.whois.cymru.com]
[v4.whois.cymru.com]
AS      | IP              | AS Name
137     | 141.250.40.34   | ASGARR GARR Italian academic..
```

Figure 2 – Example query to the IP to ASN mapping service by Team Cymru.

Once the bootstrapping node is identified, the actual node discovery phase begins. The quick and efficient discovery algorithm we present below is then used by each GIST node during its whole operating time to find and update information and metrics about all other GIST nodes present in the network. The discovery algorithm is gossip-based: all nodes periodically communicate with a randomly-selected neighbor and exchange (bounded) peering information in order to improve the quality of their own GIST nodes set. This approach, while requiring the definition of new GIST messages other than the ones already present for path-coupled signaling, is simple, and achieves fast and robust convergence as we will demonstrate in the next section. The addition of a distributed node discovery process in GIST has been necessary because, in most NSIS networks, signaling-aware nodes are scattered around and form a sparse overlay network without any warranty of node proximity. By using our gossip-based discovery procedure, every GIST node in the network eventually makes contact with other GIST nodes and can acquire a set of useful information, such as the set of supported NSLPs, the IP or GIST distance and so on. All this information forms the knowledge of the NSIS network of the node and is ultimately used to route and manage signaling messages that use the new epidemic MRM.

*1) Gossip-based node discovery messages*

In order to implement our new gossip-based GIST node discovery process, we defined three new GIST messages: Rumor, Rumor-Response and Rumor-Ack. In the following we describe the content of our new GIST messages and how they are actually used to discovery all NSIS-capable nodes.

As specified in [5], all GIST messages begin with a common header, followed by a sequence of objects in type-length-value format. The common header includes a version number, message type and size, and the ID of the destination NSLP (NSLPID).The common header also includes a GIST hop count to prevent infinite message looping and various control flags. In our case the common header does not contain any new information with respect to the standard GIST common header, the only difference is that there is a null NSLPID value specified in the appropriate field. This is because our discovery process is related to GIST only, and not to any particular signaling application running as NSLP.

Following the common header, and the optional NAT-Traversal-Object, a MRI object is included. In our implementation, we chose the path-coupled MRM as the method used to route discovery messages to other GIST nodes. In particular, we tested different types of encapsulation and message processing behaviors that use path-coupled MRM, and the results are shown in the next section. After the path-coupled MRI, all three messages include: the Session ID (SID), which is a randomly chosen 128 bit identifier that is used to relate requests to responses, and the Network Layer Information (NLI), which carries information about the network layer attributes of the node sending the message. This includes a peer identity and IP address for the sending node. It also includes IP TTL information to allow the IP hop count between GIST peers to be measured and reported, and a validity time for the GIST routing state. Finally, the Rumor and the Rumor-Response messages also include a Supported-NSLPs object, which contains the list of the supported NSLPID of the sending node, and optionally the Node-List object, which carries information about a list of nodes that are known to the sending node. The last two objects have been designed and implemented from scratch as a GIST extension to support our new GIST node discovery process.

*2) Gossip-based node discovery message exchange*

The three messages introduced in the previous subsection are then exchanged by each GIST node during its whole operating time in order to acquire and maintain information about the status of the NSIS network. The way the three gossip messages are exchanged has been designed to be very similar to the way the GIST protocol handles the three-way handshake (Query – Response – Confirm) used to create and refresh a routing state between two adjacent nodes. In the following, all GIST nodes belonging to the same network (i.e. the same AS) and discoverable by other GIST nodes are called GIST *peers*. By discovering a GIST node, we mean the ability to learn about the presence of a GIST node in the network and to acquire its GIST peer identity, supported NSLP types, IP address and a set of useful metrics, depending on the discovery approach used. This information is then used to drive the subsequent epidemic signaling dissemination.

We define three different approaches to the GIST node discovery problem depending on the way messages are exchanged by peers and on the way new nodes are discovered and added to the peer's network knowledge. The different approaches are shown in Figure 3. All three approaches are gossip-based, and specifically they belong to the background data dissemination protocols class, in which nodes periodically gossip about information associated to them thereby constantly improving the set of nodes they know. In all three approaches, the first message exchange happens with the GIST bootstrapping node or tracker. The message exchange order is fixed. The sequence consists of a Rumor message towards the destination host, then a Rumor-Response back to the source host and finally a Rumor-Ack to the destination. The messages are always sent in UDP mode and there is no control over missing responses or ack. Although the protocol is not round based at the global level, it is often convenient to refer to cycles of the protocol execution. We define a cycle to be an interval of Δ time units and it is used to estimate both the convergence speed and the communication cost.

Each GIST node executes the protocol whose pseudo-code is reported in Figure 4. Any given view contains the information about a set of nodes. The method MERGE is a set operation in the sense that it keeps at most one descriptor for each node. Parameter *m* denotes the message size as measured in the number of node descriptors that the message can hold (example in Figure 3 uses *m*=1). The method SELECTPEER selects a random sample among the known nodes, while RANDOMIZE is used to shuffle the list of known nodes without taking into account the node passed as first argument. Finally SENDACK and RECVACK are used to manage the Response-Ack messages.

After the first message exchange with the bootstrapping node, the operation is periodically repeated following the three different approaches listed below and illustrated in Figure 3.

In the first approach, called Q-mode and denoted by the number 1 in Figure 3, the periodic exchange of messages uses the GIST Q-mode encapsulation for the Registration messages. This means that the message is intercepted by on-path GIST routers at GIST distance 1 and processed. The message is not further forwarded, thus it is not possible to directly communicate with GIST nodes with GIST distance larger than 1. The subsequent Response and Ack messages are sent using normal GIST D-mode encapsulation within UDP datagrams. Thus, this scheme allows measuring the latency and IP hops only with respect to nodes at distance equal to 1 GIST hop.

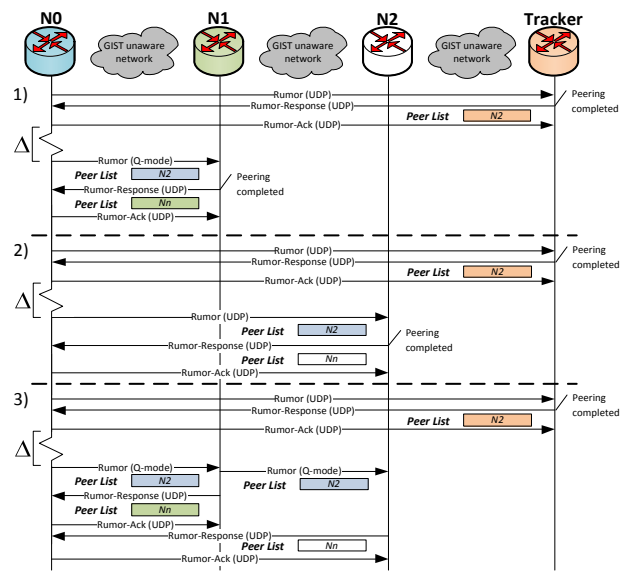

Figure 3 – Message exchange for the three different GIST node discovery alternatives.

In the second approach, called UDP-mode and denoted by the number 2 in Figure 3, all messages are exchanged using the GIST D-mode encapsulation. This means that the gossiping messages are always processed by the destination node selected by the source. In this approach, it is not possible to calculate the number of GIST hops between two peers.

In the third approach, called Q-full and denoted by the number 3 in Figure 3, the Registration messages are exchanged using the GIST Q-mode encapsulation as in the Q-mode approach. In this case, when a message is intercepted is processed as usual, but also forwarded further on-path towards the destination set by the source node. This means that all GIST nodes on-path are involved in the discovery process. From the above description, it is clear that only the third approach is able to provide, for each peer, the number of GIST/IP hops, and a (coarse) estimation of the latency.

After the network has been fully discovered, the protocol does not need to run anymore at full speed and therefore the cycle length Δ can be relaxed. Clearly, detecting global convergence is difficult and expensive: what we need is a simple local mechanism that can slow down the protocol at all nodes independently. We implemented a mechanism in which each node monitors its own local view of the network. If no changes (i.e., node additions or deletions) are observed for a specified period of time (idle), it relaxes the value of Δ or even suspends its active thread. If a view change occurs when a node is suspended (due to an incoming message initiated by another node that is newly joined or that it is communicating network changes), the node switches again to the active state, and restores the initial value of Δ. In this way also node mobility is supported, and GIST (normally limited to routers and other in-network devices) could also be integrated into end hosts.

```
1: loop
2:   wait(Δ)
3:   p ← selectPeer(view)
4:   buffer ← merge(view, {myDescriptor})
5:   buffer ← randomize(p, buffer)
6:   send first m entries of buffer to p
7:   receive buffer_p from p
8:   view ← merge(buffer_p, view)
9:   sendAck()
```
(a) active thread

```
1: loop
2:   receive buffer_q from q
3:   buffer ← merge(view, {myDescriptor})
4:   buffer ← randomize(q, buffer)
5:   send first m entries of buffer to q
6:   view ← merge(buffer_q, view)
7:   recvAck()
```
(b) passive thread

Figure 4 – The gossip-based discovery protocol.

As regards the scalability of our discovery process in ASs with a large number of GIST nodes, we made sure that the rumor message could not traverse more than a preconfigured amount of GIST or IP nodes. Also, upon sending a rumor-response message, each node can decide which nodes can be shared with the inquiring node (currently implemented using a simple IP netmask mechanism). Finally, we use a policy to decide if storing the information about a new peer based on the information relevant to the peer itself, e.g. distance in terms of IP or GIST hops, shorter than the TTL on the rumor message. In this way, each NSIS node will maintain the visibility only of GIST nodes within a pre-defined boundary.

*A. Epidemic MRM extension*

In the following, we present our epidemic MRM extension used to disseminate signaling messages within an AS. In this paper, we define and adapt to GIST three different signaling dissemination modes inspired by real issues in the area of the network and service management. Please note that the underlying IP routing is clearly independent of the GIST operation and it is carried out according to classic IP routing strategies, such as min hop routing.

- Bubble (Figure 5.a): the signaling is disseminated from *X* up to all GIST nodes with a distance equal or less than *r* from *X* according to the metric.

- Balloon (Figure 5.b): the signaling is initiated by *X*, and it is disseminated to all GIST nodes within a distance *r* from target node *Y* according to the metric.

- Hose (Figure 5.c): the signaling is initiated by *X*, and it is disseminated to all GIST nodes within a distance *r* from any GIST node in the path to the target node *Y*.

Basically, the three different modes respond to different needs in the area of network and service management. For instance, considering the NSLP proposed for the NetServ-based management architecture illustrated in [22], the bubble mode allows disseminating signaling messages for deploying monitoring modules around the source. The balloon allows a similar process around destination. As for the hose mode, it can be used to disseminate signaling messages for searching and deploying media relay/processing modules close to the IP path in mobile networks (see also [24]).

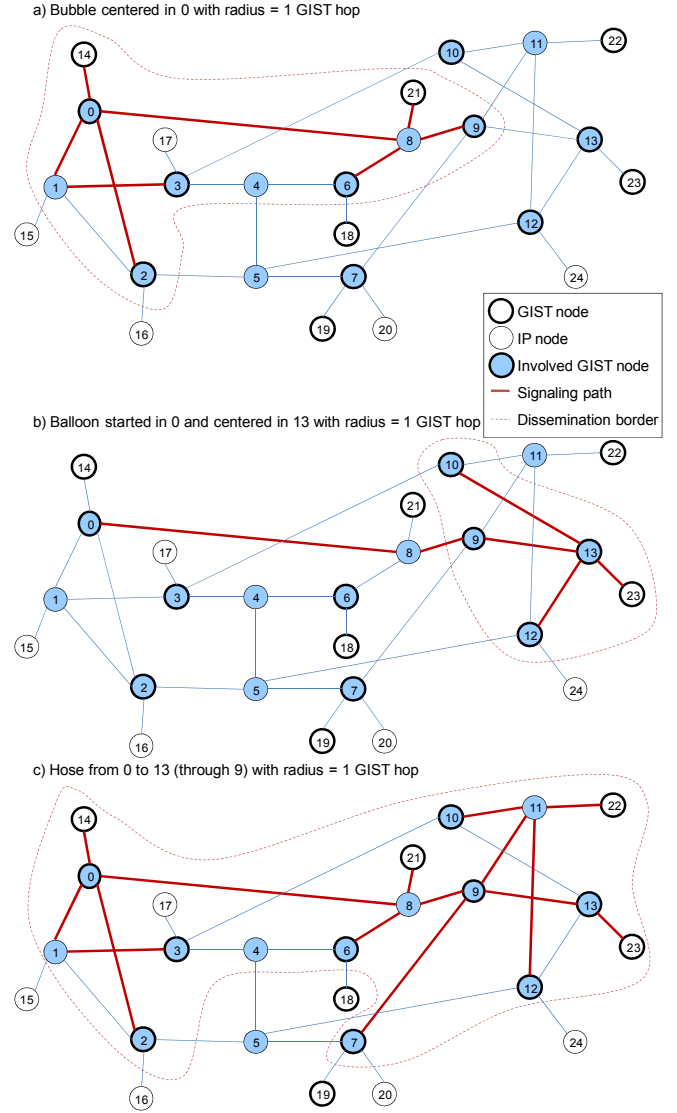

Figure 5 – Epidemic GIST message dissemination modes: a) bubble, b) balloon, and c) hose on the network topology created using GENI.

According to the information collected during the discovery phase, it is possible to define three different types of metric: i) number of GIST hops, ii) number of IP hops, and iii) latency. In fact, at the end of the gossip-based discovery, each GIST node knows the number of GIST/IP hops and has an estimation of the latency towards any other GIST node in the AS. In the example reported in Figure 5, we used as metric the number of GIST hops, and the metric value is set equals to 1 GIST hop.

The proposed epidemic MRI is based on the loose coupled MRM, using a MRM-ID set to one of the value reserved for experimental use [5]. We included an Epidemic Type field, defining one of the modes illustrated in Figure 5, and a Metric field in TLV format, used to compute the radius of the dissemination area. It is worth noting that in the bubble dissemination mode, the destination address is clearly not meaningful and it is set equal to 0.0.0.0 by the sending NSLP.

Operatively, all the three dissemination processes are based on the bubble mode. In fact, the balloon can be easily implemented by triggering a bubble on the destination GIST node, and the hose can be implemented by triggering a bubble on each GIST node on the path between the source and the destination. Thus, now we focus on how the dissemination in bubble mode is carried out. In the following, we use as the metric/distance the number of GIST hops, so for the sake of simplicity we will omit to indicate the metric type in the notation. In order to change the metric, only small modifications have to be performed. Let us define:

- $A$={the set of GIST nodes in the AS}
- $d_X(K)$ is the distance from node $X$ to node $K$.
- $N_X^K$ is the set of all the GIST nodes in the path $X$ to $K$.
- $D_X(r)=\{Y \in A \mid d_X(K)=r\}$ is the set of GIST nodes such that their distance from $X$ is equal to $r$.
- $S_X(r) = \{Y \in A \mid d_X(Y) \leq r\} = \bigcup_{l=1}^{r} D_X(l)$ is the set of GIST nodes within a circle of radius $r$ centered in $X$

Basically, disseminating signaling messages within a bubble of radius $r$ from the node $X$ means to deliver such messages to all nodes belonging to the set $S_X(r)$. A first solution, which we call *Simple Unicast*, consists of sending unicast GIST messages to *all* these destinations, known at the end of the discovery phase, separately. However, this scheme is highly inefficient, since a number of messages will be delivered over the same links a number of times. In fact, in order to reach a GIST node with distance larger than 1 (e.g. with distance equal to 3) from $X$, it is necessary to cross the GIST nodes on the same path with GIST distance 1 and 2, which are known from the discovery phase using the third approach (Q-full), in addition to a number of (unknown for $X$) IP routers. In fact, the vector $N_X^K$ is known at the end of the discovery phase.

Thus, it is possible to exploit this information to make the dissemination process more efficient. To this aim, we observe that it is enough to send a separate signaling message only to the destinations belonging to the set $G_X(r)$ to complete the bubble dissemination process (*GIST Unicast*). Figure 6 reports the pseudo-code of the algorithm used to populate $G_X(r)$.

```
1: G_X(r)=∅
2: I_X(r)=S_X(r)
3: for i in r..1
4:     G_X(r)=G_X(r) ⋃ {D_X(i) ⋂ I_X(i)}
5:     I_X(i)=I_X(i) − ⋃_{Y∈{D_X(i)∩I_X(i)}} {N_X^Y ⋂ I_X(i)}
6: end
```

Figure 6 – Pseudo-code for GIST destination selection algorithm.

In this paper we explicitly avoid considering randomized scheme for signaling message dissemination, since these schemes may exhibit very large delivery times (see e.g. the blind broadcasting protocol in [23]).

## IV. TESTBED AND PERFORMANCE EVALUATION

We have implemented the proposed solution in GENI [6], the well-known platform for network experimentations. The network topology we used is inspired to the famous NSFnet one. Figure 5 highlights GIST and non-GIST nodes, distinguished in host and routers. Each GIST node was equipped with the NetServ [7][22] NSLP to test the epidemic dissemination of in-network services.

As for the bootstrap phase, Figure 7.a reports the convergence time (average value and 95th percentile), measured in number of cycles, as a function of the different discovery algorithms presented in section III.A.2). Each cycle has a fixed duration equal to 10 seconds. The best performing algorithm is the third one (Q-full), since it exploits the native GIST interception capabilities together with randomized peer selection, typical of gossip-based solution. Thus, having enhanced the discovery process with the capability of intercepting packets at the IP layer allows improving performance with respect to pure application layer solutions, with an improvement equal to about 30% of saved time. We have repeated the test for node 0 and node 9 as tracker. Since node 9 is the one with the lowest average distance from all other GIST nodes, it exhibits a slightly better performance. Finally, we remark that using only Q-mode for sending Rumor packets is not successful, since the true discovery phase is limited to neighboring nodes, whereas other GIST nodes are found slowly and with an incomplete set of information, since they do not directly communicate with the source GIST node.

Figure 7.b reports the overhead, measured in number of messages exchanged by GIST nodes (average value and 95th percentile), as a function of the discovery algorithms. Again, the Q-full performs better than the others.

Finally, as for message delivery schemes, for space limitation in Figure 7.c we show the performance of the bubble mode only. We compared the *Simple Unicast*, the *GIST Unicast*, and another approach resembling the expansion phase of the echo-pattern [28] called *Overlay Broadcast* (see also [23]). The performance metric used to measure the overhead of the dissemination approach to implement the bubble is the number of used link in the topology of Figure 5, averaged over all possible GIST senders. It is evident that, for an increasing size of the bubble, reported in abscissa in terms of number of GIST hops, the *GIST Unicast* definitely outperforms other approaches, whereas the broadcast-style approach, applicable once the bootstrap phase ends, is the worst.

## V. RELATED WORK

There are several existing alternative approaches that try to extend NSIS. The Generic Ambient Network Signaling (GANS) [9] protocol aims at supporting signaling sessions that may have no direct relation to any underlying data flow. GANS is designed as a backwards-compatible extension of GIST and supports dynamic interworking and convergence of heterogeneous networks. However its role-based abstract addressing scheme and its associated DEEP discovery protocol do not support multiple signaling destinations. Another alternative is GISP, General Internet Service Protocol [10], which is defined as a NLTP according to the architectural framework defined by the NSIS working group (Figure 1). The main difference with GIST is that GISP implements the signaling and transport services itself while GIST provides a common messaging layer which relies on existing transport and security protocols. GISP also does not support different MRMs specification, resulting in a less flexible protocol architecture.

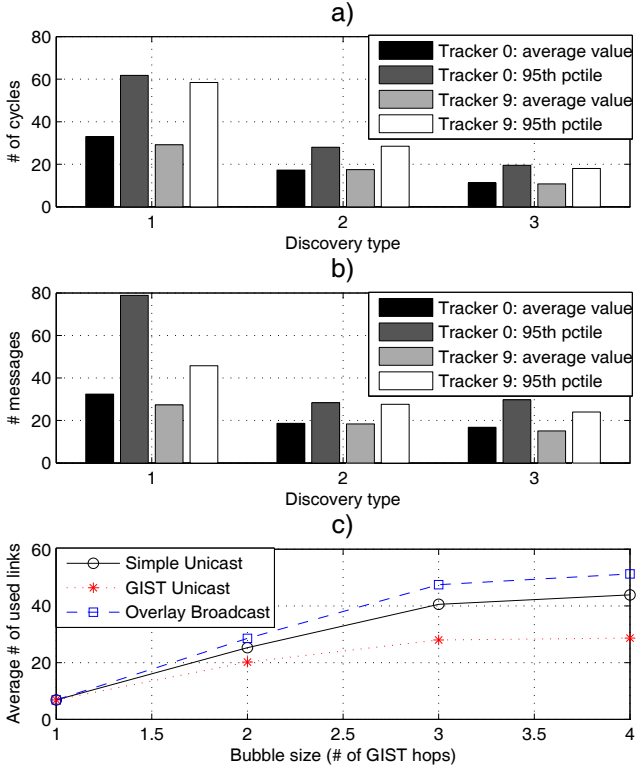

Figure 7 – Performance of the proposed solution: (a) number of gossip cycles and (b) number of messages needed to discover GIST nodes in the test network, and (c) number of links used to disseminate messages in bubble mode.

Our approach has been inspired by the lack of gossip-based, or epidemic [11][12], dissemination methods in GIST and their usefulness, especially in scenarios where, to perform the neighbor discovery function, flooding is not possible since target nodes form a sparse network (as in the current case of NSIS nodes) and there is need for a bounded worst-case load and/or delivery time. Among the many alternative protocols that use a gossip-based scheme to propagate events or to synchronize replicated data, we particularly make reference to the T-Man protocol [13] for our bootstrap process and to the organization of the NTP pool [14] for our tracker discovery process using DNS queries.

An alternative way to perform bootstrapping operations is using an anycast address [15]. This implies that a network administrator must allocate and maintain the special anycast address for all GIST nodes in its managed network and also configure all GIST nodes to use it, including the update of routers' routing tables. Even though this solution is the simplest to implement, it is inefficient because it would not support all the different signaling message routing methods that we propose. Only the bootstrap process could leverage from the new addressing scheme, but only if there is an anycast address associated to all GIST nodes inside each AS.

SNMP [16] broadcast discovery could also be used to find GIST nodes in a network, but its usage is currently discouraged for security reasons whereas GIST supports the usage of additional authorization and authentication schemes. Also, not all devices inside a network are SNMP-managed whereas NSIS, and by extension GIST, being a generic signaling suite, can run on both routers and end-hosts alike.

Other GIST peer discovery alternatives could have been OSPF, with its additional router capabilities advertising [17] and its link local signaling [18], and SLP (Service Location Protocol) [19]. Both these approaches are not applicable to our case. The former works only on routers and modifications to the routing daemon are often unwelcome and discouraged by network operators. The latter works only in local area networks, where it adopts Service Agents to inform User Agents about available services through an extensive usage of multicast/broadcast traffic. In order to scale to larger networks (e.g. enterprise networks), it needs to employ centralized repositories for advertised services named Directory Agents, which present all the problems of centralized architectures [24].

Regarding the dissemination mode, it is necessary to cite the echo pattern [28], defined in the context of active networks to discover functions available in surrounding nodes. In fact, this scheme, in its expansion phase, resembles the bubble and the balloon dissemination modes. Please note that the echo pattern cannot be used in the node discovery phase in our case (sparse network), since it can work only if the all network nodes are able to understand and relay signaling messages, in order to propagate the "wave" upon reaching the termination condition (e.g. TTL expiration), or if they already have information about next hop in the overlay topology. In addition, it is worth citing also the path-directed search pattern introduced in [24], which resembles our proposed hose dissemination model, although the old approach has not been designed to support deployment in sparse networks.

VI. CONCLUSION AND FUTURE WORK

In this paper, we have presented an extension of the NSIS protocol family to support an epidemic-style signaling message dissemination. This extension has been implemented at the GIST layer, in order to be usable by all NSLPs. We have shown that initial results are very promising, in terms of both convergence time, overhead, and message delivery efficiency.

Future work will address a complete performance evaluation. In addition, we are implementing a multicast-like dissemination algorithm, to further improve the dissemination efficiency. Finally, for balloon and hose dissemination modes, we have to define additional methods to cope with non-GIST destinations.